\documentclass[usenatbib,letterpaper]{mn2e}

\usepackage{graphicx}
\usepackage{appendix}
\usepackage{psfrag}
\usepackage{amsmath,amssymb} 
\usepackage{threeparttable}
\usepackage{float}
\usepackage{subfigure}
\usepackage{afterpage}
\numberwithin{equation}{section} 
\title[ALFALFA H{\sc i} Data Stacking II]{ALFALFA H{\sc i} Data Stacking 
II. \\H{\sc i} content of the host galaxies of AGN.} 
\author[S. Fabello]{Silvia Fabello$^{1}$\thanks{fabello@MPA-Garching.MPG.DE}, Guinevere Kauffmann$^{1}$, Barbara
  Catinella$^{1}$,  Riccardo Giovanelli$^{2}$, \newauthor
 Martha P. Haynes$^{2}$, Timothy M. Heckman$^{3}$, David Schiminovich$^{4}$
 \\
$^{1}$Max-Planck Institut f\"{u}r Astrophysik, D-85741 Garching, Germany\\
$^{2}$Center for Radiophysics and Space Research, Cornell University, Ithaca, NY 14853, USA\\
$^{3}$Department of Physics and Astronomy, The Johns Hopkins
  University, Baltimore, MD 21218, USA\\
$^{4}$Department of Astronomy, Columbia University, New York, NY 10027, USA
}

\begin{document}          

\def\deg{$^{\circ} $}
\newcommand{\mone}{$^{-1}$}
\newcommand{\Ha}{$\rm H\alpha$}
\newcommand{\Hb}{$\rm H\beta$}
\newcommand{\hi}{{H{\sc i}}}
\newcommand{\hii}{{H{\sc ii}}}
\newcommand{\nii}{[N{\sc ii}]}
\newcommand{\oiii}{{[O{\sc iii}}]}
\newcommand{\Msun}{M$_\odot$}
\newcommand{\Lsun}{L$_\odot$}
\newcommand{\Mhi}{M$_{\rm HI}$}
\newcommand{\Mst}{M$_\star$}
\newcommand{\must}{$\mu_\star$}
\newcommand{\col}{NUV$-r$}
\newcommand{\cix}{$C$}
\newcommand{\df}{$D_n(4000)$}

\maketitle

\label{firstpage}
\begin{abstract}
We use a stacking technique to measure the average {\hi} content of a
volume-limited sample of 1871 AGN host galaxies
from a parent sample of galaxies selected from the SDSS and
GALEX imaging surveys  with stellar masses greater
than $10^{10} 
M_{\odot}$ and redshifts in the range $0.025<z<0.05$. 
{\hi} data are available from the Arecibo Legacy Fast ALFA (ALFALFA) survey.
In previous work, we found that the {\hi} gas fraction in galaxies correlates most strongly
with the combination of optical/UV colour and 
stellar surface mass density. We therefore  build a
control sample of non-AGN matched to the AGN hosts in these two properties.
We study trends in {\hi} gas mass fraction ({\Mhi}/{\Mst}, where
{\Mst} is the stellar mass) as a function
of black hole accretion rate indicator L{\oiii}/M$_{BH}$. 
We find no significant difference in {\hi} content between AGN and control
samples at all values of black hole accretion rate probed by 
the galaxies in our sample. This indicates that AGN do not influence
the large-scale gaseous properties of galaxies in the local Universe. 
We have studied the variation in {\hi} mass fraction with black hole
accretion rate in the blue and red galaxy populations. In the blue
population, the {\hi} gas fraction is independent of accretion rate,
indicating that accretion is not sensitive to the properties of the
interstellar medium of the galaxy on large scales. However, in the  
red population accretion rate and gas fraction do correlate.
The measured gas fractions in this population are not too different
from the ones expected from a stellar mass loss origin, implying that 
the fuel supply in the red AGN population could be a mixture 
of mass loss from stars and gas present in disks. 

\end{abstract}


\section*{Introduction}
Observations have shown that most low redshift massive galaxies
contain a black hole (BH) in their bulges \citep{Korm04} and that in some cases,  the
black hole may be actively accreting.
The mass of the black hole is strongly coupled
with the mass and velocity dispersion  of the bulge of its  host galaxy
\citep{Korm95,Mag98,Gebhardt00, Ferrarese00, Tremaine02}. 
Whether or not energy output by  accreting black holes can influence the evolution
and properties of their host galaxies is  still a matter of debate. 
In some theoretical models of galaxy evolution, (radio) AGN are assumed to heat the external gas reservoir
in massive dark matter halos \citep[eg.][]{Bower06, catt06, Croton06,
  DeLucia07, Som08}, thereby quenching the growth of the most massive galaxies. It is believed that this process may
solve the over-cooling problem in simulated galaxy clusters
\citep[eg.][]{Chur01, Reyn01, Quilis01, Ruszkowski02, Sijacki06, Tey10}. Another favored 
evolutionary scenario is that  the energy output by the AGN actually
drives gas out of the galaxy itself, depleting its interstellar medium, and shutting down
star formation \citep[e.g.][]{DiMatt05, Hopk06}.  

Observationally, the situation is still very unclear.
In nearby galaxies \citep[e.g.][]{Kauff03b, Heck04,Schaw07} and
galaxies at intermediate redshifts \citep[e.g.][]{Nandra07, Silver08},
the  host galaxies of AGN are found to occupy 
the ``green'' part of the  colour - magnitude 
diagram, i.e. their stellar populations are intermediate in age between
the blue, actively star-forming spiral galaxies, and the  gas-poor, quiescent
early-type population. One interpretation of these results is that there is an  
evolutionary sequence driven by  nuclear activity: i.e. AGN are  triggered
in galaxies on the blue
sequence, gas in these systems is heated, and  galaxies  move
onto the red sequence as star formation shuts down 
\citep{Schaw09}. However, this interpretation is certainly not unique. 
At a fixed black hole mass, AGN tend to be found preferentially 
in galaxies with younger stellar
populations (and hence more gas). The presence of a gaseous fuel 
supply may simply be
a necessary condition for both star formation and  accretion 
onto the black hole \citep{Kauff03b, Kauff09}. One way to ascertain whether 
AGN are responsible for terminating 
the growth of massive galaxies, is to establish a direct link between the presence of
an active nucleus and the {\em gas content} of such galaxies.
  
There have only  been a few {\em systematic} studies of the 
{\hi} content of the host galaxies of 
moderately large samples of AGN. 
\cite{Ho08} 
conducted a survey of neutral gas in a representative sample of
154 moderate to high luminosity nearby Type I AGN.
Their results appear to challenge the feedback scenario because AGN 
are found to be  at  least as gas rich as
quiescent objects of the same morphological type. 
\cite{Hugh09} carried out a study of {\hi} in a volume-limited sample of galaxies
in different environments around the Virgo cluster, trying to
disentangle between the effect of AGN and environment on star
formation. 
In their sample, the incidence of AGN peaks in the ``green valley'', 
but galaxies with quenched star formation are found to be mainly HI
deficient galaxies located in high-density 
environments. No direct connection between  
AGN activity and quenching of star formation was observed, at odds
with the previous cited studies.   

In this paper we measure the average {\hi} content of a sample of
$\sim$2000 AGN host galaxies, using the stacking technique
developed in \citet[][hereafter Paper I]{Fab10}. 
In \citet[][]{gass01} and Paper I we showed that the gas content of  
galaxies with stellar masses greater than
$10^{10} M_{\odot}$ is, to first order, most tightly
correlated with optical/UV ({\col}) colour and stellar surface mass density.
We build control samples that are very closely matched
to our AGN sample in the ({\col})/$\mu_*$ plane, and we
look for differences in {\hi} mass fraction {\Mhi}/{\Mst}
between AGN and control galaxies. In section 1 we describe
the sample selection and the data we use; in section 2 we compare in
bins of nuclear properties the gas content of AGN and inactive
counterparts. 

\section{Samples}
Our targets are drawn from the volume-limited, homogeneous
\textit{sample A} defined in Paper I.  To summarize, this sample
consists of 4726 galaxies with
stellar masses greater than $10^{10} M_{\odot}$ and redshifts in the
range $0.025<z<0.05$ selected from the SDSS main spectroscopic sample
and the projected GALEX Medium Imaging Survey \citep[MIS;][]{galex}.
HI data are available from the Arecibo Legacy Fast ALFA Survey
\citep[ALFALFA;][]{alfalfa}. Among our targets, 23\% have
been detected by ALFALFA. We refer the reader to Paper I for
further details on the sample selection and for a description of the
stacking method used to evaluate average {\hi} gas fractions.     

From this main sample, we extract a sample of AGN
($\S\,$\ref{sample_agn}), and a control sample of galaxies where there is 
no evidence of accretion onto the central 
black hole ($\S\,$\ref{sample_cntr}). In the next section, we  introduce 
the parameters we use for both the analysis and the sample selection.

\subsection {Galaxy Parameters}\label{pars}
The optical/UV parameters we use are derived from SDSS spectrum
measurements\footnote{See http://www.mpa-garching.mpg.de/SDSS/DR7/ and
http://cas.sdss.org./dr7/en/tools/search/sql.asp} and reprocessed
GALEX UV photometry \citep[][Paper I]{JWang10}. From
there we use: the {\col} colour measured for the whole galaxy
(corrected for Galactic extinction only); the stellar mass surface
density {\must}={\Mst}/$(2\pi R^2_{50,z})$,where $R_{50,z}$ is the Petrosian radius
containing 50\% of the flux in $z$-band in units of kpc; the central
velocity dispersion $\sigma$.  
In addition we use the 4000{\AA} break measured from the SDSS
3 arcsec-diameter fiber spectrum, {\df}. As discussed in  
\citet{Kauff03a}, {\df} can serve as a measure of the  age of 
the stellar population of galaxies, although there is a secondary
dependence on metallicity.
The galaxies in our sample are located at low redshifts
($0.025<z<0.05$), and in 72\% of our sample the 3 arcsec fiber covers
less than 20\% of the Petrosian radius enclosing 90\% of the
total light of the galaxy. 
The 4000{\AA} break values should thus be regarded as a measure of 
stellar population age for    
the inner nuclear regions of the galaxies, while the
{\col} colour is a proxy for the specific star formation rate SFR (sSFR,
i.e. SFR per unit stellar mass). 

Finally, we extract the following emission fluxes: {\Ha}, {\Hb},
{\nii}, {\oiii} (in units of $10^{-17}\,$erg$\cdot$s$^{-1}$cm$^{-2}$), 
which we  use to classify the nuclear activity of our
galaxies by means of the  diagnostic diagram first defined by
\citet[][hereafter BPT]{BPT81}.  
Because the fiber covers the inner region of the galaxy,
the AGN properties should not be significantly contaminated by outer star forming
regions \citep[see][]{Kew06}.  
 
As discussed in \citet{Kauff03b}, the
luminosity of the {\oiii}$\lambda$5007 line should be
a reasonably reliable tracer of black hole accretion rate. Because
it is a high excitation line, it is  
less contaminated by emission from {\hii} regions than
other lines such as {\Ha}.
We correct our line fluxes for dust extinction using the Balmer decrement,
following eq. (3) in \citet{Wild07}, and then convert the {\oiii}
fluxes into luminosities.  

Finally, we evaluate an Eddington ratio for the AGN, defined as: 
\begin{eqnarray*}
  \frac{L_{bol}}{L_{EDD}}\,\propto\,\frac{L_{bol}}{M_{BH}}\,\propto\,
  \frac{L[OIII]}{\sigma^4}
\end{eqnarray*}
where $L_{bol}$ is the bolometric luminosity, and $M_{BH}$ the central
black hole mass. The Eddington luminosity $L_{EDD}$ scales with
$M_{BH}$, which we estimate using the stellar  velocity
dispersion $\sigma$ measured from the
fiber spectrum: $M_{BH}\propto \sigma^4$. $L_{bol}$ scales with
L{\oiii} \citep[see][]{Heck04}, so to  first order
the ratio L{\oiii}/$\sigma^4$ can be used as a proxy for the accretion
rate. For a precise estimate of the Eddington parameter we refer the
reader to \citet{Heck04} and \citet{Kauff09}.

\begin{figure}
\includegraphics[width=8.2cm]{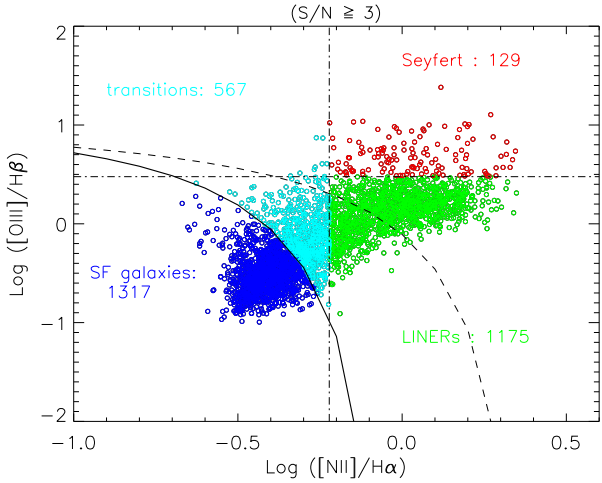}
\caption{BPT diagram for galaxies in our sample  with S/N$\ge$3 for  the four
  emission lines {\Ha}, {\Hb}, {\oiii} and {\nii}. 
  The solid curve we use to demarcate the boundary between AGN and ``normal''
  star-forming (SF) galaxies is
  from \citet{Kauff03b}; the dashed line shows the demarcation boundary for
  ``pure'' AGN  from  \citet{Kew01}. Objects are
  colour-coded according to their nuclear properties as labeled in the
  diagram.}\label{fig01} 
\end{figure} 

\subsection{The AGN sample}\label{sample_agn}
We extract a sample of AGN from \textit{sample A} using the
BPT diagnostic diagram, which allows to distinguish Type 2 AGN from
star-forming galaxies. 
We consider the ratio {\oiii}/{\Hb} versus {\nii}/{\Ha} for all
galaxies with signal-to-noise ratio S/N$\,\ge\,$3 in \emph{all} 
four line measurements.  
Following \citet{Kauff03b}, AGN are defined as those galaxies with
\begin{eqnarray*}
\log\, (\mathrm{[OIII]/H\beta})\, \ge \, 0.61/\{\log
(\mathrm{[NII]/H\alpha})-0.05\}+1.3,
\end{eqnarray*}
We obtain a sample of 1871 active galaxies, which can be further subdivided into
Seyferts, LINERs (which make up the bulk of our sample) or ``transition''
as indicated in Figure \ref{fig01}.

We note that \citet{Kew01} suggested a more stringent cut 
to separate galaxies where the nuclear emission is almost
completely excited by emission from the gas accreting onto the black holes,
rather than from star-forming regions
(shown as a dashed line in Figure \ref{fig01}). With this cut, we
obtain a subsample of 912 ``pure'' AGN, which is too small for our stacking
analysis. This is why we only consider the demarcation line suggested by
\citet{Kauff03b} in this paper. 

In the top panel of Figure \ref{fig02}, we plot
the AGN in the colour - stellar mass surface
density plane. 
The ``pure'' AGN defined using the \citet{Kew01} cut (dashed line in
Figure \ref{fig01}) are over-plotted as magenta points. As can be seen,
both samples span the same range of parameter space.

\subsection{The control sample}\label{sample_cntr}

\begin{figure}

\begin {tabular}{c}
\includegraphics[width=8.cm]{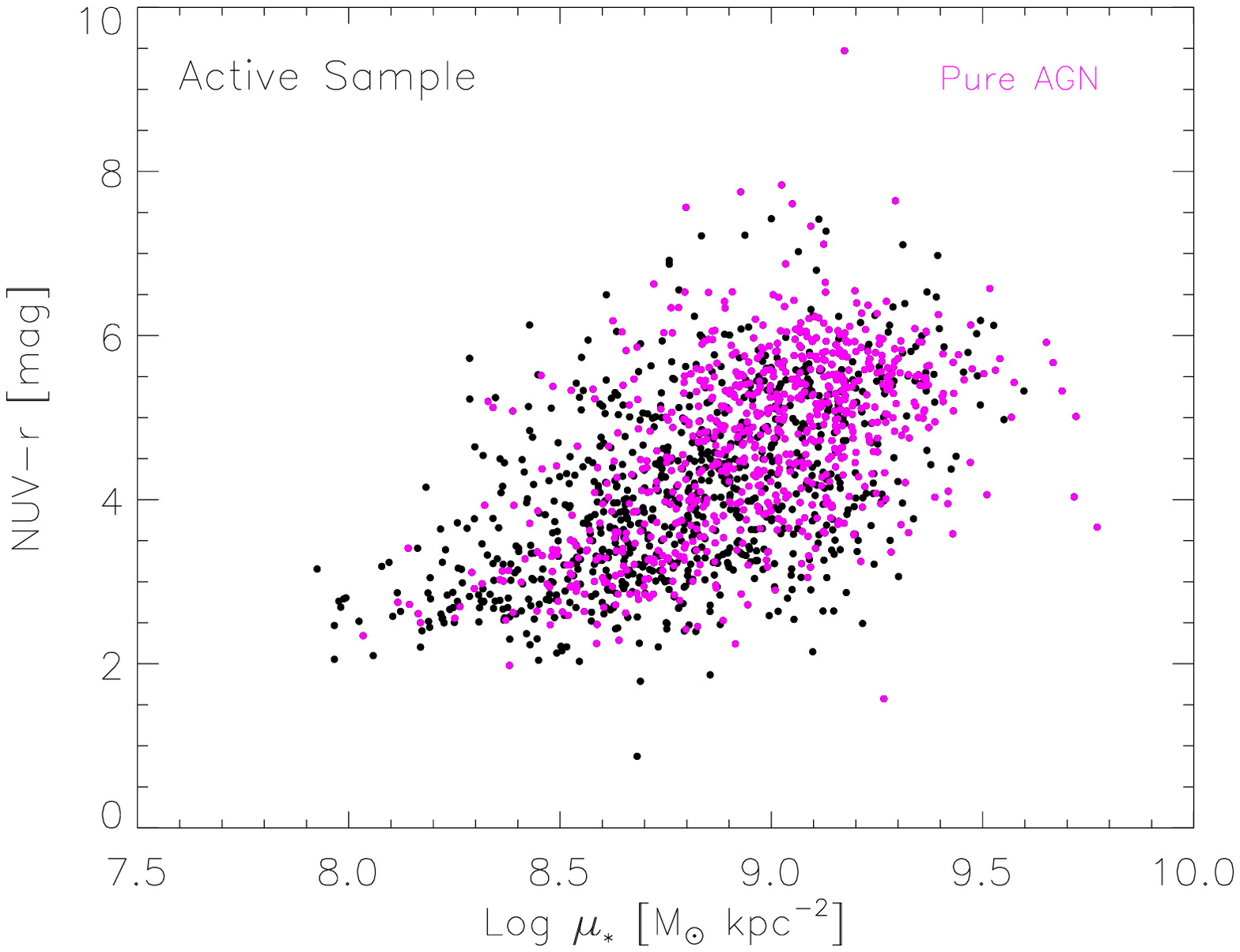}\\
\includegraphics[width=8.cm]{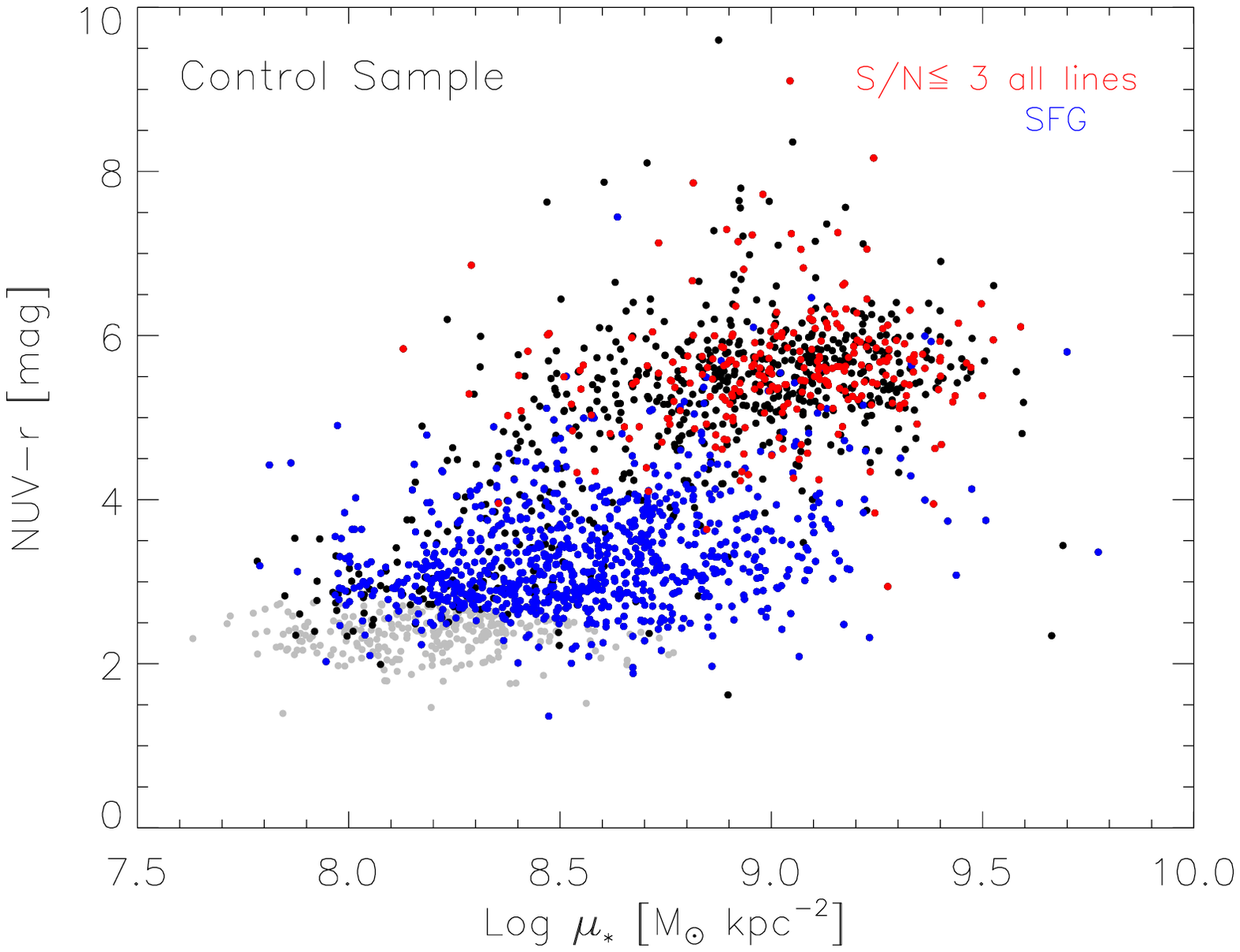}
\end {tabular}\caption{Galaxies in our samples are plotted
in the ({\col})/{\must} plane. 
\emph{Top}: Black points show the
\textit{AGN sample}. We overplot as magenta points galaxies that lie above the 
stronger cut defined by
\citet{Kew01}. \emph{Bottom}: All control pool galaxies, made by 
``inactive'' galaxies with S/N({\oiii})$<\,$3 plus star-forming objects.
Gray dots are the galaxies that are discarded 
when matching to the AGN sample (i.e. mainly very blue, low
stellar surface density objects). Black and coloured dots show the
final \textit{control sample} extracted. In particular, galaxies for which all four emission   
lines have S/N$\,\le\,$3 are indicated as red points, while the blue points show galaxies 
that are classified as star-forming.}\label{fig02}  
\end{figure} 

Our control sample is drawn from a combined sample of ``inactive'' galaxies
and star-forming systems.
Inactive galaxies are those where the  S/N of  the {\oiii} line is less than
3. We exclude 474 galaxies where the S/N of the {\oiii} line is
greater but other lines lie below this  S/N threshold, because these galaxies
cannot be accurately classified. By applying a cut in the {\oiii} line
only, our control sample may be contaminated by weak LINERs. The
contamination is anyway negligible. For our targets, less than 5\% of the galaxies with
S/N({\oiii})$\,<\,$3 have $\log
(\mathrm{[NII]/H\alpha})>\log(0.6)$ (vertical dash dotted line in Figure \ref{fig01}).\\
Inactive galaxies, together with galaxies classified as
star-forming on the BPT diagram, constitute our pool of 
2377 galaxies for which the nuclear emission is not
dominated by an AGN. The colour-{\must} plane for those
  galaxies is shown in the bottom panel of Figure
\ref{fig02}, all points. Two distinct populations are clear in this diagram: 
a red sequence of galaxies with weak or absent emission
lines, and a blue cloud consisting of star-forming 
galaxies which extends until lower {\must} with respect to the active
sample. 

We stack AGN in bins of the accretion rate proxy L{\oiii}/$\sigma^4$. For each bin, 
we want to compare the average {\hi} mass fraction with
that obtained for control 
galaxies matched in stellar mass surface density and
{\col} colour. For each AGN we then select a galaxy from the control
pool by searching for its closest neighbor in the plane defined by 
{\col} and {\must}. We match objects in  order of decreasing
{\must}, and we do not allow a  galaxy to enter the control sample more
than once. We do not have enough galaxies to build larger control
samples, but with this method we are able to reproduce  the original
{\must} and {\col} distributions of the \textit{AGN sample}. The gray
dots in Figure \ref{fig02} (bottom panel) are the objects discarded
because they do not match any AGN. The black and
coloured points show the final \textit{control sample}. In particular, blue
dots are star forming galaxies, 
while  galaxies for which none of the four emission lines
{\Ha}, {\Hb}, {\oiii} and {\nii} are detected with $S/N >3$
are overplotted as red dots.

\section{The HI gas fractions of AGN hosts}

\begin{figure*}
\begin {tabular}{cc}
\includegraphics[width=8.cm]{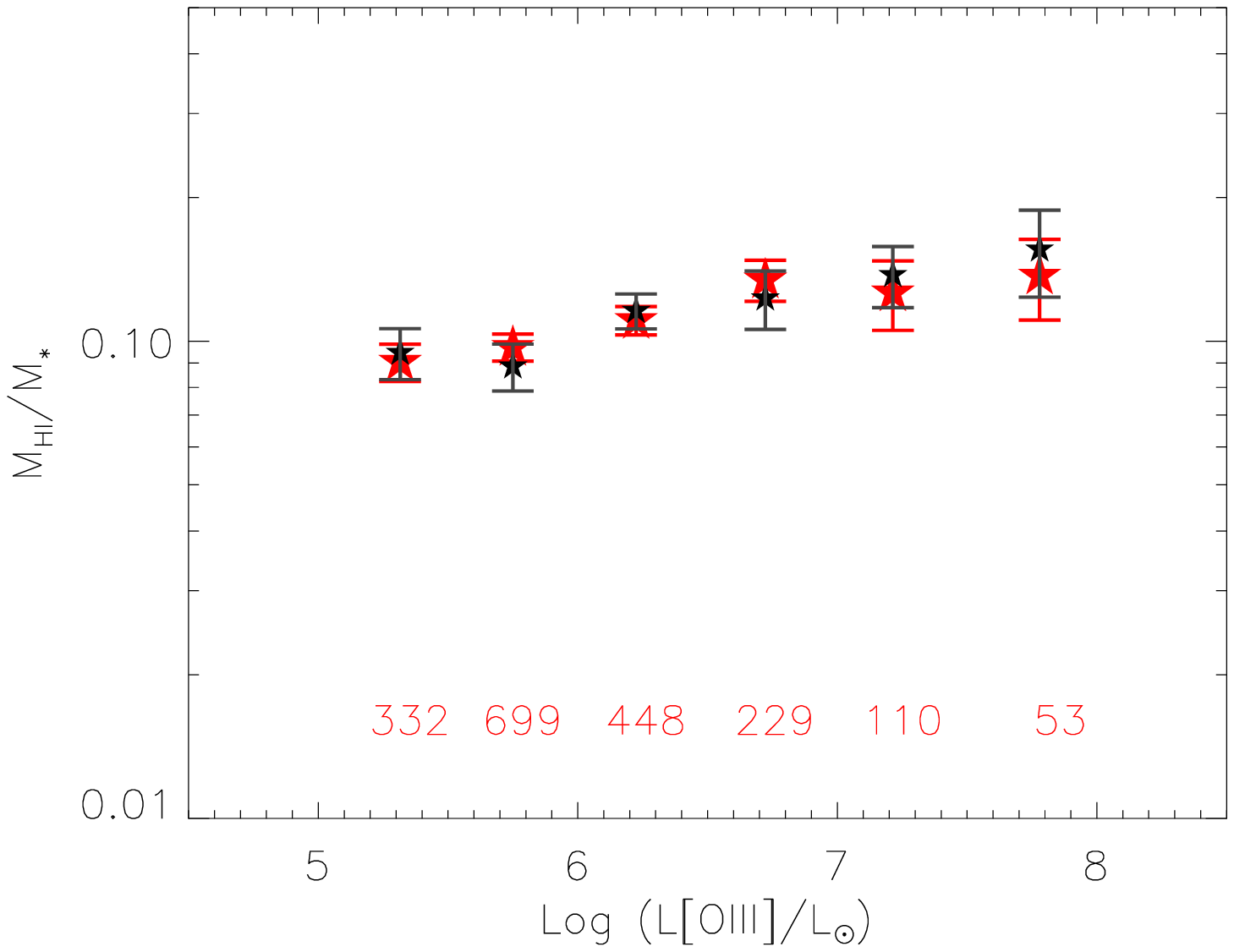} &
\includegraphics[width=8.cm]{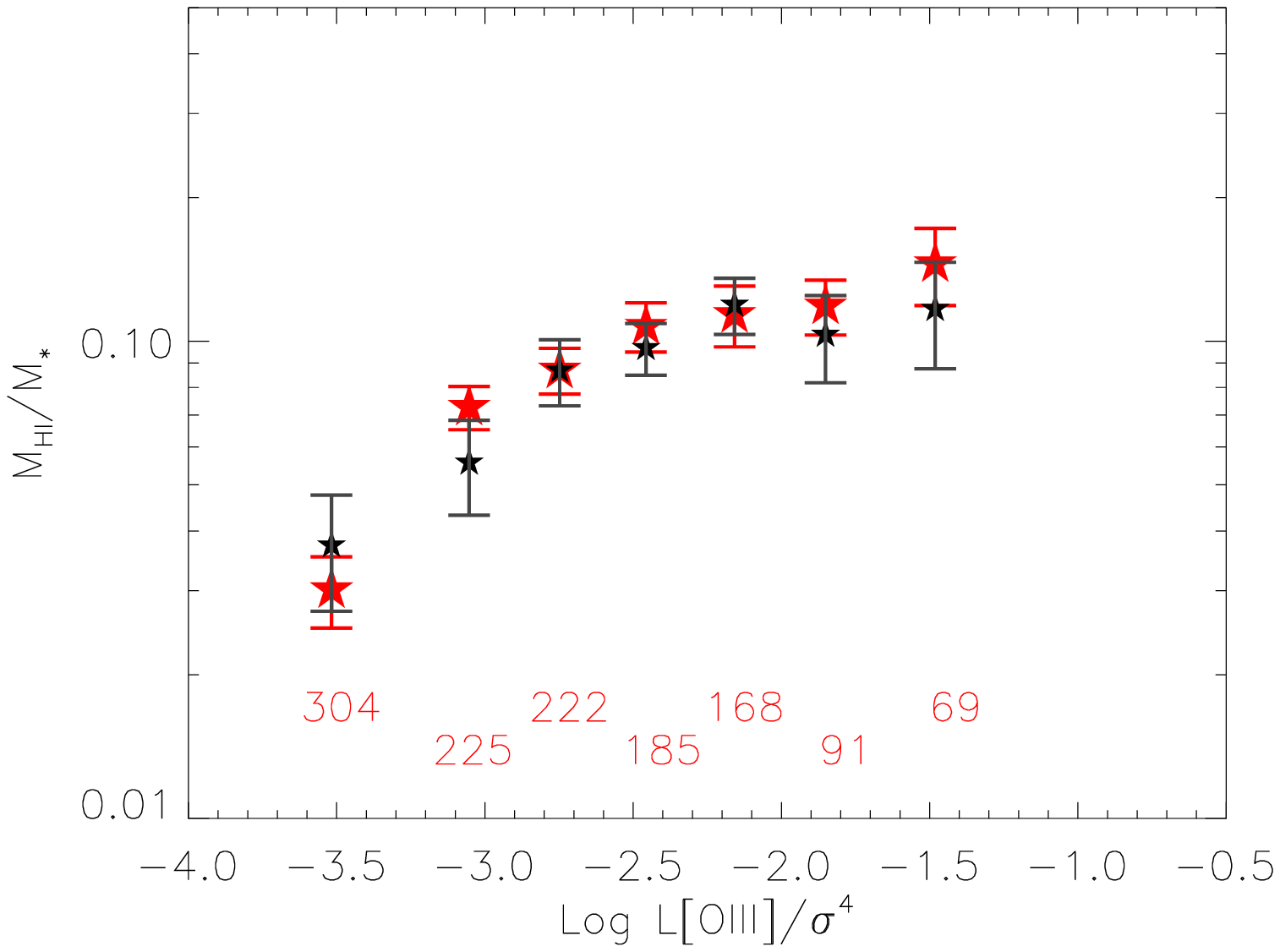} 
\end {tabular}\caption{{\hi} gas fraction as a
function of L{\oiii} (\emph{left panel}) and L{\oiii}/$\sigma^4$
(\emph{right panel}). Filled red stars are the values obtained for the
AGN hosts, while the black stars are the values obtained for
the corresponding control galaxies. 
The number of galaxies included in each stack is reported at
the bottom of the plot at the position of each bin.}\label{fig03} 
\end{figure*}

We have stacked our AGN sample in bins of {\oiii} line luminosity and
in bins of the black hole accretion rate proxy L{\oiii}/$\sigma^4$.
In Figure \ref{fig03}, we plot the average gas fraction {\Mhi}/{\Mst}  
as a function of both parameters for both the AGN sample (red) and
for the control galaxies (black).
Errors are evaluated using the bootstrap
method for a total of  200 repetitions: every time we discard
20\% of the sample and evaluate a new gas fraction; the error 
is the dispersion on the repeated measures. 

As can be seen, the average {\hi} mass fraction increases weakly as a
function of {\oiii}, from about 9\% for the weakest AGN to 
$\sim$14\% for the most luminous systems.
This nearly flat trend reflects the fact that AGN of different {\oiii}
luminosities are located in roughly the same region of the 
{\must}-({\col}) plane. The dependence of {\hi} mass fraction on black hole accretion rate 
is considerably stronger. Objects with the lowest measurable
accretion rates are very gas poor ({\Mhi}/{\Mst}$\sim$3\%), and the average gas
fraction increases as a function of L{\oiii}/$\sigma^4$ up to
a maximum value of $\sim$14\%. We likewise find that  
galaxies with higher black hole accretion rates
tend to be bluer and have lower values of {\must}. 
We note that the trend in {\hi} mass fraction as a function of black hole accretion rate
is considerably weaker than as a function of the optical/UV colour of the
host galaxy. In Paper I,
we found that the average {\hi} mass fractions change from $\sim$55\% for the
bluest objects to $\sim$2\% for the reddest ones.  

\begin{figure*}
\begin {tabular}{cc}
\includegraphics[width=8.cm]{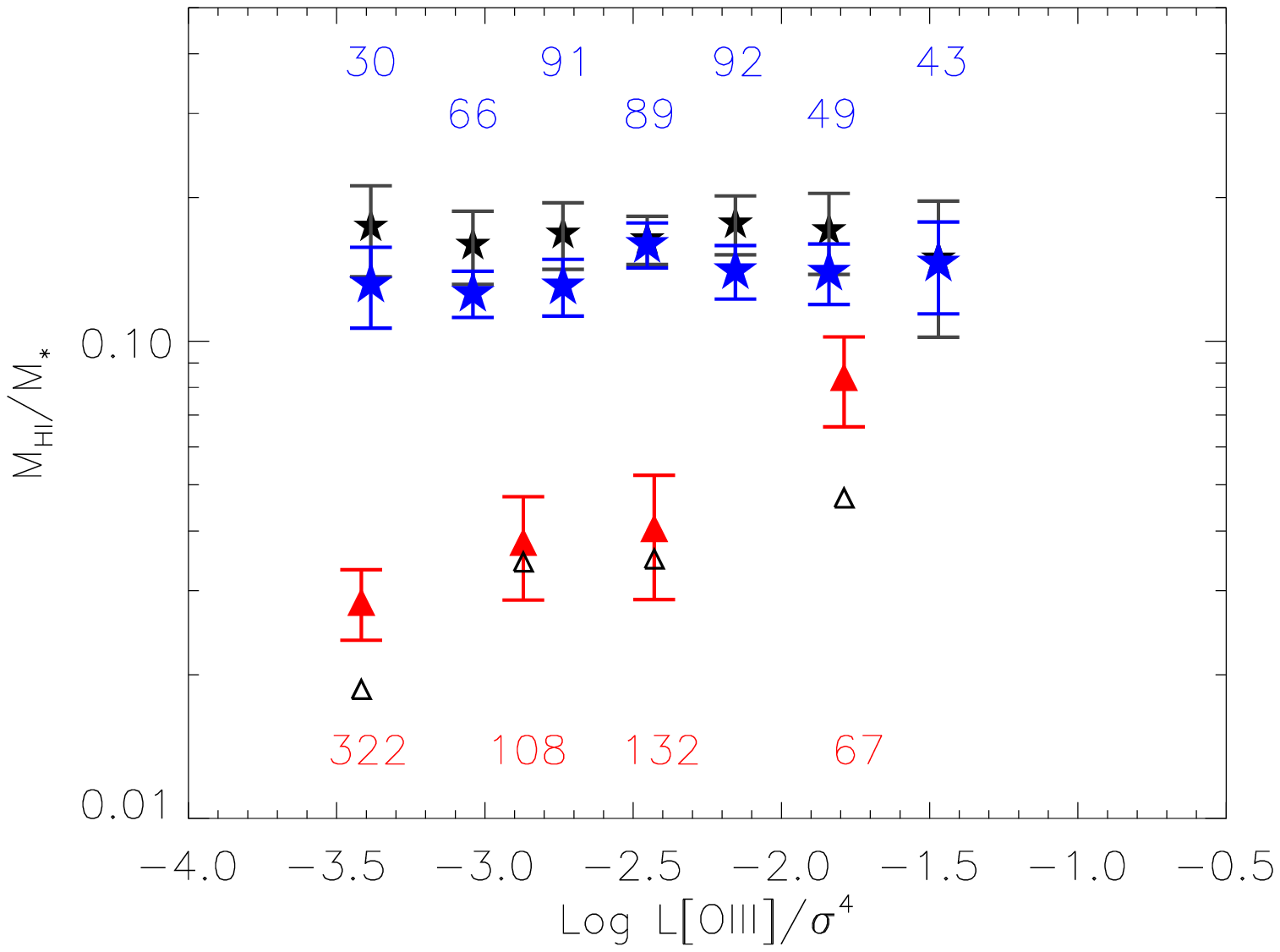} &
\includegraphics[width=8.cm]{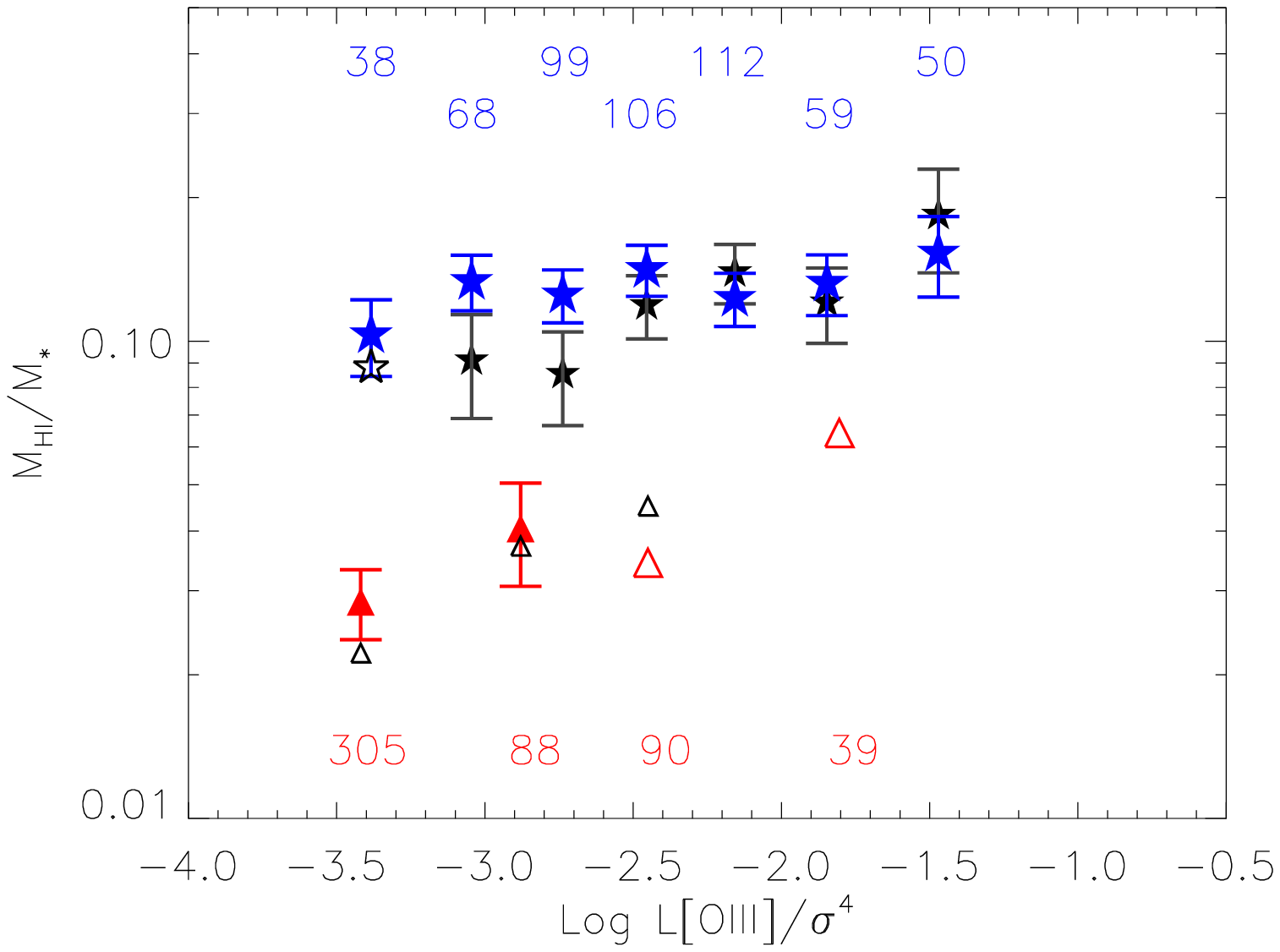}\\
\end{tabular}\caption{ {\hi} gas fraction as a function of L{\oiii}/$\sigma^4$
for galaxies and AGN hosts, split into two bins of star formation properties. Coloured symbols show results 
for galaxies  with active galactic nuclei, while black symbols show results
obtained for the corresponding control samples. The empty
symbols denote measurements that are upper limits. The number of
objects stacked in each bin is reported.  \emph{Left panel}: the trend
with gas fraction is shown separately for blue galaxies (defined as {\col}$<$4.5), and red sequence ones
({\col}$>$4.5). \emph{Right panel}: galaxies are matched and divided into two bins
of specific star formation rate; blue symbols are for star forming
targets, red for quiescent ones.} \label{fig04} 
\end{figure*}

The next step is to assess whether AGN differ from control galaxies in terms
of mean {\hi} gas mass fraction. Results for the control sample are plotted as 
black stars in Figure
\ref{fig03}, at the x-axis value of the corresponding AGN bin.
Errors are evaluated as explained previously.
As can be seen, there is no difference between
the neutral gas content for the two samples.

\citet{Kauff09} showed that there appear to be two distinct
regimes of black hole growth in galaxies in the local Universe.
The first is associated with galaxies with significant ongoing star
formation; in this regime the distribution of accretion rates shows
little dependence on the central stellar population of the galaxy. 
The second regime is associated with ``passive'' galaxies, and is 
characterized by a power-law distribution of accretion rates.
There, the accretion rate does depend strongly on the age of
the central stellar population in the galaxy.

We now take a look at {\hi} gas mass fraction trends for blue and
red galaxy sub-populations. We define red sequence galaxies to
  have {\col}$\,\ge\,$4.5, and blue sequence galaxies to have
  {\col}$\,<\,$4.5. This cut is based on the split in the bimodal
  colour distribution of the control sample, visible also in Figure
  \ref{fig02} (lower panel).
Results are shown in Figure \ref{fig04} (left panel): blue stars are blue cloud
AGN, black stars are the corresponding control galaxies; red triangles
represent the red sequence AGN, while black triangles are used
for control galaxies. Symbols are empty when the measure is a
non-detection (upper limit). 
The average {\hi} gas fraction remains constant as a function of the
accretion rate parameter L{\oiii}/$\sigma^4$ for the blue population.
The increase in {\hi} gas fraction as a function of  L{\oiii}/$\sigma^4$ 
seen in Figure \ref{fig03} is in fact driven by galaxies on the
red sequence. We note that we had to increase the bin size for these
objects in order to recover sufficient signal in our stacks.
Our control samples of red and blue galaxies exhibit the same trends
with  L{\oiii}/$\sigma^4$ as the AGN, but the gas mass fractions are slightly 
higher for the blue control objects, and lower for the red control
objects. (In fact, we do not detect {\hi} in any of the red control galaxy stacks).

The {\col} colours we use are not corrected for dust
extinction. It is conceivable that AGN may be found in galaxies
with more dust than average, so we should ascertain whether
using uncorrected colours to
create  matched  control samples will
bias our results.     
Reliable estimates of the specific star formation rates of our
galaxies may  be obtained by fitting spectral energy distribution models to
the 7-band UV and SDSS data, and calibrating the attenuation as function of
{\col} colour using a reference sample with direct measurements 
of the UV through far-IR spectral energy distribution 
from a combination of GALEX, SDSS
and Spitzer data. Our methodology  is explained in
detail in Saintonge et al. (2011, MNRAS, submitted). 

We repeat our
stacking exercise, this time matching AGN hosts to control galaxies
using {\must} and specific star formation rate rather than
{\must} and NUV-r colour. The new
results are shown in the right panel of Figure \ref{fig04}.
Red symbols denote objects with low specific star formation rates
($\log\,$SFR/{\Mst}$\,<\,$-11.0), while blue symbols are for
more strongly star-forming galaxies
($\log\,$SFR/{\Mst}$\,>\,$-11.0). Once again, this division has been
 made based on the  bimodal distribution of sSFR in  our sample. 
We see the same trends as before, with no significant difference in HI content
between the blue AGN hosts and control galaxies.
The same conclusion holds for red AGN hosts, although we note that
the two highest accretion rate bins yield non-detections. This is because
many of these galaxies shift over to the blue sequence after dust
corrections are made. 
 
\begin{figure}
\includegraphics[width=8.cm]{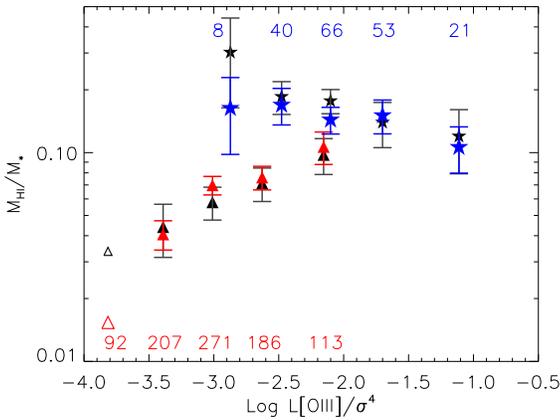} \caption{{\hi}
    gas fraction as a function of L{\oiii}/$\sigma^4$, for galaxies 
  divided into two subsamples according to 4000{\AA}
  break strength: red triangles are galaxies with
  {\df}$>$1.7 and blue stars with {\df}$<$1.6. Symbols are as in Figure \ref{fig05}. }  \label{fig05} 
\end{figure}

We note that the amount of accretion onto the central supermassive
black hole in a galaxy is more likely to be sensitive to the properties
of the interstellar medium in the central parts of the galaxy rather
then globally. We now investigate what happens if we split our AGN sample
into ``red'' and ``blue'' subsamples using {\df}, which is measured within the
3 arcsec diameter SDSS fiber aperture, and remeasure gas fractions
for these two new samples, plus corresponding control galaxies. 
In this case, blue galaxies have {\df}$<$1.6 and red galaxies
have {\df}$>$1.7 (this roughly separates the peaks of the bimodal
distribution of 4000{\AA} break strengths).  
In Figure \ref{fig05} we show our results. 
AGN hosts exhibit no significant differences in their gas 
content with respect to non-AGN in the same {\df} range 
matched by global colour and {\must}. Also the differences in the red population
seen in Figure \ref{fig04} has now vanished, given that now we
detect the same signal for both AGN and control galaxies.  We
note that the two bins of {\df} clearly separate in terms of the trend
in gas fraction as a function of the Eddington parameter L{\oiii}/$\sigma^4$, 
consistent with the suggestion of
\citet{Kauff09} that accretion onto the black hole is supply-limited in the
red galaxy population, but not in the blue one.

\section{Summary \& Discussion}

We have used the stacking technique developed  in a
companion paper (Paper I) to look at possible interplay between
nuclear activity and gas content in massive nearby galaxies. Starting
from a volume-limited, homogeneous sample of massive galaxies we have
selected a sample of AGN, and a control sample of inactive galaxies
matched by {\must} and {\col}. We then stacked their {\hi} data
in bins of {\oiii} line luminosity scaled by
black hole mass to assess whether the gas content of AGN hosts
differs from that of the control galaxies.  

 Our main result is that we 
do not find any significant difference between the {\hi} gas
content of AGN hosts and that of control galaxies. This conclusion holds at all
values of the Eddington parameter proxy L{\oiii}/$\sigma^4$ probed by the
galaxies in our sample.\\
 If converted into a proper Eddington ratio, our sample would
  cover the range $\log (L/L_{Edd})\,\sim\,$-4 to -2. We are studying the
  relatively low accretion rate regime, and that could be a reason why we do not
  see any sign of AGN feedback. We would need data over a larger
  volume to stack a significant number of AGN with very high
  L/L$_{Edd}$.  Still, up to the highest accretion rates we have studied,
  the gas fraction shows no hint of correlation with either
  L{\oiii}/$\sigma^4$ (blue population) or AGN presence.

These results are also in agreement with
\citet{Ho08}. As discussed in their paper, the lack of any difference may
be due to the fact that the {\hi} traces gas on very large scales, but accreting 
black holes only heat gas close to the centers of the galaxies.

In some simulations of galaxy formation, AGN feedback is an extremely violent process  
that can drive out much of the interstellar medium from galaxies,
truncating ongoing star formation and reddening their  stellar
populations \citep{DiMatt05}. 
In this scenario, AGN hosts would be  expected to be gas
deficient. Catastrophic global
feedback processes are clearly not occurring in the majority
of AGN in the local Universe. We cannot yet ascertain that AGN feedback
is important in the centers of galaxies. In future work, we plan to
examine trends in molecular gas fractions in AGN hosts using
data from the COLD GASS survey (Saintonge et al 2011). The molecular
gas is generally more centrally concentrated that the atomic gas
and may be more sensitive to feedback processes occurring in the
bulge of the galaxy. 

Another interesting result of our analysis is the scaling of the mean {\hi}
gas fraction of AGN host galaxies with accretion rate onto
the black hole, as traced by L{\oiii}/$\sigma^4$.
In the blue galaxy population, the accretion rate onto the black hole
and the {\hi} gas fraction of the galaxy are independent.
However, in the red galaxy population, these two quantities do track
each other. \citet{Kauff09} hypothesized that in blue galaxies, 
the black hole regulates its own growth at a rate that does not 
further depend on the properties of the interstellar medium.
In red galaxies, they deduced that the decrease in the accretion 
rate onto black holes in old galaxies was  consistent 
with population synthesis model predictions of the decline in 
stellar mass loss rates as a function of mean stellar age. 

Because we now have measured {\hi} mass fractions for the red population,
we can check this in more detail. The average global ({\col})/($g-r$) colours of the
host galaxies vary from 4.56/0.74 for the stack with the highest accretion rate
in Figure \ref{fig05} to 5.54/0.80 for the stack with the lowest accretion rate.
As described in \citet{Kauff09} we use population synthesis models
\citep[e.g.,][]{BC03,Maraston05} to calculate  mass loss
rates as a function of  colour,  
assuming simple exponential declining star formation histories for our galaxies.
If the {\hi} content of red sequence AGN came only from mass loss, we
would expect a 0.3 dex change in the average $\log\,$({\Mhi}/{\Mst})
between AGN stacks with the lowest and highest accretion rates in our
sample. Figure \ref{fig05} shows a 0.5 dex change  
in this quantity, which is somewhat larger than predicted. 
However, we note that the average stellar surface mass densities of the
host galaxies in the stacks decrease by a factor of 3 between AGN with
the highest and lowest accretion rates (likewise the concentration
index decreases from 3.15 to 2.79), so structural properties are
not constant across the population. This may  imply that 
the fuel supply in the red AGN population is in fact a mixture 
of mass loss from stars and gas present in disks. 

\section*{Acknowledgments}
We thank the many members of the ALFALFA team who have contributed to
the acquisition and processing of the ALFALFA dataset over the last six years.

RG and MPH are supported by NSF grant AST-0607007 and by a grant from the Brinson
Foundation.

The Arecibo Observatory is part of the National Astronomy and
Ionosphere Center, which is operated by Cornell University under a
cooperative agreement with the National Science Foundation.

GALEX is a NASA Small Explorer, launched in 2003 April. We gratefully
acknowledge NASA's support for construction, operation and science
analysis for the GALEX mission, developed in cooperation with the
Centre National d'Etudes Spatiales (CNES) of France and the Korean
Ministry of Science and Technology.

Funding for the SDSS and SDSS-II has been provided by the Alfred
P. Sloan Foundation, the Participating Institutions, the National
Science Foundation, the U.S. Department of Energy, the National
Aeronautics and Space Administration, the Japanese Monbukagakusho, the
Max Planck Society and the Higher Education Funding Council for
England. The SDSS web site is http://www.sdss.org/. 

The SDSS is managed by the Astrophysical Research Consortium for the
Participating Institutions. The Participating Institutions are the
American Museum of Natural History, Astrophysical Institute Potsdam,
University of Basel, University of Cambridge, Case Western Reserve
University, University of Chicago, Drexel University, Fermilab, the
Institute for Advanced Study, the Japan Participation Group, Johns
Hopkins University, the Joint Institute for Nuclear Astrophysics, the
Kavli Institute for Particle Astrophysics and Cosmology, the Korean
Scientist Group, the Chinese Academy of Sciences (LAMOST), Los Alamos
National Laboratory, the Max-Planck-Institute for Astronomy (MPIA),
the Max-Planck-Institute for Astrophysics (MPA), New Mexico State
University, Ohio State University, University of Pittsburgh,
University of Portsmouth, Princeton University, the United States
Naval Observatory and the University of Washington.


\end{document}